\documentclass[]{pasj01}

\begin{document} 
\Received{2016/03/08}
\Accepted{2016/09/11}

\title{Machine-learning Selection of Optical Transients
in Subaru/Hyper Suprime-Cam Survey}

\author{Mikio \textsc{Morii}\altaffilmark{1}}
\author{Shiro \textsc{Ikeda}\altaffilmark{1}}
\author{Nozomu \textsc{Tominaga}\altaffilmark{2,3}}
\author{Masaomi \textsc{Tanaka}\altaffilmark{4,3}}
\author{Tomoki \textsc{Morokuma}\altaffilmark{5,3}}
\author{Katsuhiko \textsc{Ishiguro}\altaffilmark{6}}
\author{Junji \textsc{Yamato}\altaffilmark{6}}
\author{Naonori \textsc{Ueda}\altaffilmark{6}}
\author{Naotaka \textsc{Suzuki}\altaffilmark{3}}
\author{Naoki \textsc{Yasuda}\altaffilmark{3}}
\author{Naoki \textsc{Yoshida}\altaffilmark{7,3}}
\altaffiltext{1}{Research Center for Statistical Machine Learning,
The Institute of Statistical Mathematics,
10-3 Midori-cho, Tachikawa, Tokyo 190-8562, Japan}
\altaffiltext{2}{Department of Physics, Faculty of Science and
Engineering, Konan University,
8-9-1 Okamoto, Kobe, Hyogo 658-8501, Japan}
\altaffiltext{3}{Kavli Institute for the Physics and Mathematics
of the Universe (WPI), The University of Tokyo,
5-1-5 Kashiwanoha, Kashiwa, Chiba 277-8583, Japan}
\altaffiltext{4}{National Astronomical Observatory of Japan,
2-21-1 Ohsawa, Mitaka, Tokyo 188-8588, Japan}
\altaffiltext{5}{Institute of Astronomy, The University of Tokyo,
2-21-1 Ohsawa, Mitaka, Tokyo 181-0015, Japan}
\altaffiltext{6}{NTT Communication Science Laboratories,
2-4, Hikaridai, Seika-cho, Keihanna Science City, Kyoto 619-0237, Japan}
\altaffiltext{7}{Department of Physics, The University of Tokyo,
7-3-1 Hongo, Bunkyo-ku, Tokyo 113-0033, Japan}
\email{morii@ism.ac.jp}



\maketitle

\begin{abstract}
We present  an application of machine-learning (ML) techniques
to  source selection in  the optical transient survey data
with Hyper Suprime-Cam (HSC) on the Subaru telescope.
Our goal is to select real transient events accurately and in a timely manner
out of a large number of false candidates,
obtained with the standard difference-imaging method.
We have developed the transient selector  which is based on  majority voting of three ML machines
of AUC Boosting, Random Forest, and Deep Neural Network.
We applied it to our observing runs of Subaru-HSC in 2015 May and August, and
proved it to be efficient  in selecting  optical transients.
The false positive rate was 1.0\% at the true positive rate of 90\%
in the magnitude range of 22.0--25.0 mag for the former data.
For the latter run, we successfully detected and reported ten candidates of
supernovae within the same day as the observation.
From these runs, we learned the following lessons:
(1) the training using artificial objects is effective
in filtering out false candidates, especially for faint objects,
and (2) combination of ML by majority voting is advantageous.
\end{abstract}

\section{Introduction}

The 8.2-m Subaru telescope has  been running
a 300-night Strategic Survey Program (SSP)
over 5 years since 2014 March\footnote{http://www.naoj.org/Projects/HSC/surveyplan.html},
in order to elucidate the mystery of dark matter and dark energy
as well as the evolution of galaxies.
The survey utilizes Hyper Suprime-Cam (HSC; \cite{Miyazaki+2012})
with a wide field of view of 1.77 square degrees.
In 2016--18, we observe two ultra deep fields, COSMOS and SXDS,
for six-month each.
The observations will be performed around the new moon
in each month, with a typical cadence of 3--4 days.
Note that the HSC is not installed on Subaru for two weeks
around the full moon.
Among other transient surveys,
the Subaru HSC/SSP survey is the deepest for this survey area
($1.77 \times 2$ square degrees), and thus,
will provide the unique dataset for transients.
For example, the HSC/SSP survey will
triple the number of type-Ia supernovae (SNe) beyond redshift $z>1$
and will also discover a few tens of superluminous SNe at $z>1$.
 
Difference imaging is the standard method to search for optical transient
objects, and so we use it in this study.
We define a transient object as the one that  appears only
in the image at the later epoch (newer image),
but not in the one at the earlier epoch (reference image or template image), out of the two images taken at different epochs.
After the standard data reduction is  made,
two images are astrometrically aligned, and
the reference image is subtracted from the newer image
by matching point spread functions (PSFs) of the two images.
The source-finding algorithm is applied to the difference image, and  the detected sources are
the candidates of transients
(section \ref{sec: Data Analysis and Feature Extraction}).

\begin{figure}
 \begin{center}
   \includegraphics[width=7cm, angle=0]{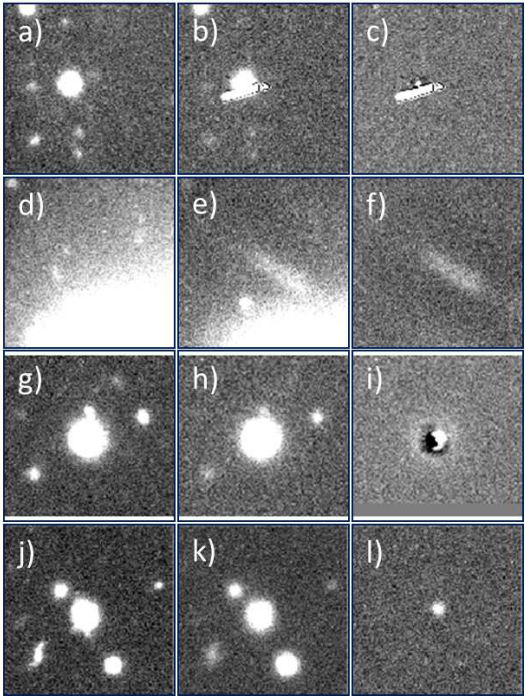}    
 \end{center}
\caption{Examples of real and bogus objects obtained  with Subaru-HSC.
The left,  middle, and right columns show
 the reference, new, and difference images, respectively.
The first, second, and third rows show 
the cosmic ray (a--c), ghost near a bright star (d--f), and
inaccurate image convolution or astrometric alignment (g--i),
respectively.
The bottom row shows a real transient located in a galaxy (j--l).
}\label{fig:bogus_sample}
\end{figure}

In an ideal situation, all the sources detected in difference images
 would be  transient/variable astronomical sources, such as
supernovae, variable stars, moving objects, and so on.
In reality, however, they also include artifacts
(see panels a--i in figure \ref{fig:bogus_sample}),
such as cosmic-ray events, spikes around bright stars,
and residuals related to inaccurate image convolution or astrometric alignment.
These artifacts are  present in every
optical survey project \citep{Bailey+2007, Bloom+2012, Brink+2013}.
Hereafter, we call them ``bogus'' \citep{Bloom+2012}.

In the HSC/SSP survey, not only a few hundred transients, including SNe,
but also $\sim 10^5 - 10^6$ bogus objects are expected to be detected each night.
After the scheduled 300 nights, the number of candidates of transients, 
 real and bogus combined, will reach $\sim 10^8$, 
which is  well qualified as Big Data.
We need to filter out bogus objects  to select SNe and other real transients.
 Processing of filtering must be  performed swiftly in order to increase
 the chances of new findings in an early phase of
transient phenomena.

 The primary method to  distinguish transient/bogus objects is, traditionally, visual inspection by human checkers,
as many surveys have  been adopting.
However, the expected size of our data  is so big that human checkers would
not be able to go through all the data in a reasonable time.
We have decided to introduce 
machine-learning (ML) techniques to select real transients.

In the filtering process,
we should not miss  real objects,
while a vast number of bogus objects are filtered out.
Throughout the development,
we try to minimize the false positive rate (FPR),
while we maintain the true positive rate (TPR) of 90 \% or larger 
(namely the false negative rate, FNR $< 10\%$).

We performed two HSC observations in 2015 May and August
\citep{tominaga15atel1, tominaga15atel2}, aiming to detect short transients
with a time scale of a few hours to a few days
(e.g. \citet{Tanaka+2016}; \citet{Morokuma+2016}).
An example of such short transients is an optical flash at the time of
shock breakout of a supernova,
of which the time-variance would be detectable
during an observation for a single night.
We use three kinds of ML methods, 
AUC Boosting, Random Forests (RF), and Deep Neural Network (DNN), both individually and in a combined way.
We verify the performance of these machines
by making the receiver operation characteristic (ROC) curves.

Conditions of observations (the noise and seeing)
vary every night.  Hence, the ML classifiers must be robust 
against the change of environment.
We use normalized features to reduce the influence of the variation.
To validate the performance of our method, we show the
ROC curves of our machine trained 
with the data on one-night observation,
applied to the data on the other night observation.
The results show the proposed classifier is robust.

In optical surveys, ML techniques
have been introduced by \citet{Bailey+2007}
for the data from the Nearby Supernova Factory. 
They applied Boosted Decision Trees, RF, and
Support Vector Machines (SVM) and succeeded  in
reducing the number of bogus candidates by a factor of ten.
The Palomar Transient Factory team \citep{Bloom+2012}
used RF, and  achieved the TPR of 92.3\% 
at the FPR of 1\%
\citep{Brink+2013}.
The Pan-STARRS1 Medium Deep Survey used
Artificial Neural Network, SVM and RF,  and  achieved
TPR of 90\% at FPR of 1\% \citep{Wright+2015}.
\citet{Goldstein+2015} applied the RF for Dark Energy Survey Supernova program (DES-SN),
and reduced the number of transient candidates  by a factor of 13.4, which were then fed to human scanning.
\citet{du Buisson+2015} applied the RF, $k$-nearest neighbor, and the SkyNet artificial
neural net algorithm, using features trained from eigen-image analysis for
the Sloan Digital Sky Survey supernova survey.

The rest of this paper is  structured as follows.
In section~\ref{sec: Data Analysis and Feature Extraction},
we explain the HSC data reduction and feature
extractions. 
In section~\ref{sec: ML}, we introduce three machine learning methods we used. 
In section~\ref{sec: experiment}, the applications to the actual Subaru data are presented, and then 
we discuss the result of real vs bogus segregation in section~\ref{sec: Discussion and Conclusion}.

Prior to the forthcoming HSC/SSP Transient survey,
this paper  will provide a `path finder'
to identify real  astronomical
objects and to demonstrate the power of  machine learning.

\section{Data Analysis and Feature Extraction}
\label{sec: Data Analysis and Feature Extraction}

\begin{longtable}{ll}
\caption{List of the features}
\label{table:feature}
\hline\hline
\multicolumn{1}{l}{Feature variable} &
\multicolumn{1}{l}{Description} \\
\hline
\endfirsthead
mag     & Magnitude \\
magerr  & Error of magnitude \\
elongation\_norm & Elongation normalized by nearby stars \\
fwhm\_norm     & FWHM normalized by nearby stars \\
significance\_abs & Significance obtained  with the PSF fit \\
residual       & Residual of PSF fit \\
psffit\_sigma\_ratio & See text \\
psffit\_peak\_ratio  & See text \\
frac\_det & See text \\
density & Number of objects around the target within a square with 120 $\times$ 120 pixels
centered  at the target \\
density\_good & ``Density'' after weak screening \\
bapsf & Elongation of nearby stars \\
psf & FWHM of nearby stars \\
\hline
\end{longtable}

We describe the flow of HSC data reduction and feature extraction for machine learning.
The pipeline processing,
using the on-site data analysis system \citep{furusawa11},
and  then transient finding \citep{tominaga15}, are performed immediately after the data acquisition.

The HSC data are reduced  with the HSC pipeline
(version 3.6.1), which  has been developed based on the LSST pipeline
\citep{ivezic08,axelrod10}.
First, the pipeline  performs the standard reduction,
such as bias subtraction and flat fielding.
Then, astrometric and photometric calibrations are made,
using the Sloan Digital Sky Survey DR8 data \citep{aihara11}.
Finally, mosaic solution is derived and images are warped
to align astrometrically 
in preparation for co-adding (if needed) and image subtraction
for difference-imaging at the next stage.

For image subtraction we use the HSC pipeline.
The algorithm to match the PSFs of the two images
is the same  as that by \citet{alard98} and \citet{alard00}
\footnote{http://www2.iap.fr/users/alard/package.html},
 and adopts a position-dependent convolution kernel.
The optimal convolution kernel is derived so that 
the difference between convoluted PSFs becomes the smallest.
The algorithm has been implemented also in the HOTPANTS package
\footnote{http://www.astro.washington.edu/users/becker/v2.0/hotpants.html}.

Source detection is performed in  difference images,
using the HSC pipeline instead of the standard tool for it,
SExtractor \citep{bertin96}, because
the former outperforms the latter 
 in reducing  bogus detections.\newline  
For each source, the features listed
in table \ref{table:feature}  are computed as follows. We fit the  image with the PSF of two-dimensional Gaussian
and subtract the best-fit  PSF from  it.
Then, we measure the residuals within 2 $\times$ FWHM radius
($\sigma_{\rm on}$) and in the surrounding region (3--4 $\times$ FWHM, $\sigma_{\rm off}$).
We define ``psffit$\_$sigma$\_$ratio'' $= \sigma_{\rm on}/\sigma_{\rm off}$;
 it is expected to be much larger than unity
when the shape of  a source is  very different from PSF.
We also define ``psffit$\_$peak$\_$ratio'',
the ratio of the actual peak of the source
to the peak of the best-fit Gaussian in order to quantify the degree of deviation of
 the source-image shape  from the PSF.

To  obtain features sensitive to mis-alignment,
we count the number of positive pixels within $3 \times$ FWHM,
 and define ``frac$\_$posi'' as  their fraction.
This feature frac$\_$posi should  be close to unity
for   good detection with a high S/N ratio.
Similarly, we count the number of negative pixels and define ``frac$\_$nega'' for its ratio to detect ``sources'' with negative counts.
To use both the cases in a consistent manner, we define
``frac$\_$det'', which is the same  as ``frac$\_$posi'' for
the positive detection  but ($1 -$ ``frac$\_$nega'') for the negative detection.
Following \citet{Bloom+2012} and \citet{Brink+2013},
we also count the number of the detected sources inside 
a 120 $\times$ 120 pixel box centered  at the source (``density'').

In the HSC pipeline,
PSF fitting for the detected sources  is
performed  with the position-dependent PSF.
We also compute the significance of  detection (significance$\_$abs)
by comparing the fitted PSF with the noise level around   a source.

In total, 13 features  are used for machine learning (table \ref{table:feature}).

\section{Methods of Machine Learning for Real-Bogus Separation}
\label{sec: ML}

\subsection{AUC Boosting}
\label{sec: AUC boosting}

Boosting is a method to classify the data
by majority-voting of weak classifiers 
\citep{Hastie-Tibshirani-Friedman_2009}.
Among various boosting methods,
we employed the AUC boosting
developed by \citet{Komori_2011},
which is trained to maximize the empirical
area under the ROC curve (AUC).
The AUC boosting classifies objects according to
a score function, where the objects with larger scores are regarded as real.
This method uses only one hyper-parameter $\lambda$ 
to control the smoothness of the score function,
which  is optimized through cross-validation.

Once the machine  has been trained, 
classification for  a new set
of feature variables will be fast.
Hence, when the trained machine is installed in 
the pipeline process of HSC,
real and bogus are classified quickly.
Note that the computation speed for the classification
is fast for the following two methods, too.

\subsection{Random Forest}

Random Forest (RF; \cite{Breiman_2001})
is an ensemble-learning method using an ensemble of decision trees.
Each decision tree is trained with a subset of training data.
Classification is based on majority-voting of the decision trees.
For a training data set $X=\{{\bf x}_1, {\bf x}_2,\ldots,{\bf x}_n\}$ 
with labels $Y=\{y_1, y_2,\ldots,y_n\}$,
($y=0$ or 1, corresponding to bogus or real, respectively),
each of $B$ training sets  $X_b$ ($b=1,...,B$) 
is selected by bootstrap sampling (i.e., randomly drawn with replacement 
from $X$). Then, the $b$-th decision tree $f_b$ is trained 
with $X_b$ with the corresponding label set $Y_b$.
When  each decision tree is trained, a subset of the features
is also randomly selected. 

RF is considered to be robust  against outliers and noise
because its majority-voting of low-correlated classifiers
decreases the variance of the model.
The meta parameters of RF are the number of trees, $B$,
and the number of subsampled features, $m_f$. 
Following the standard practice, we use cross-validation
 to evaluate the performance and  determine the appropriate $B$ for our observation data,
and set $m_f$  $=$ $\sqrt{p}$, where $p$ is the number of all the features.
We employ the RF implementation in scikit-learn\footnote{An open-source machine learning library for python, http://scikitlearn.org}.
The standard setting for the classification task is used. 

\subsection{Deep Neural Network}

Deep Neural Network (DNN) or Deep Learning \citep{LeCun15}
is the current state-of-the-art technique,
which achieves the best performance in speech and image recognitions.
The network consists of multiple layers 
of neurons with directional connections.
The neural network is trained to tune the connecting weights
between neurons and parameters of emission functions,
so that it approximates a mapping 
from the input observations (feature vector; ${\bf x} \in {\bf R}^p$)
fed into the input layer
to the output observations emitted from the output layer
(bogus or real; $y = 0, 1$).
Efficient stochastic optimization algorithms
are used to tune millions of weights and parameters.
In our case, we use a fully-connected feed-forward network.

We use the Chainer library\footnote{
It is provided by Preferred Network Inc., http://chainer.org/}.
For the parameter estimation, we use
the stochastic gradient descent (SGD) methods.
The SGD computes the noisy gradient of the objective function 
based on the mini-batch, a subset with  $M (\ll N)$,
out of   $N$ samples. 
DNNs have many meta parameters to be tuned.
We choose them via preliminary cross-validations.

\section{Experiment}
\label{sec: experiment}

We performed HSC observations on 2015 May 24 UT and August 19 UT with a high cadence
\citep{tominaga15atel1, tominaga15atel2},
aiming to detect short-timescale transients, of which the time-variance would be detectable
during an observation for a single night.
From the observational data of May, optical transients were detected and screened
by conventional visual inspection, and 48 definitely real transient objects, most likely supernovae, 
were identified (see the next sub-section).
Note that they were not used for training machines but for performance validation of the machines
(section \ref{sec: result sel SN}).
We created data set for training based on the May data (section \ref{sec: Artificial real objects})
and trained machines with them. The machines were applied to the August data
and transients events were identified (sections \ref{sec: result sel SN} and \ref{sec: quick sel SN}).

To validate the robustness of the machines against variation of environment,
we also did the reverse; i.e., we trained the machines with the August data 
and  applied them to  the May data and their
 48  definite transients.
 In the training, we added artificial  transient objects
because the number of real transient objects in the data is  much smaller than
 that of the bogus. We explain how we created the training data
in section \ref{sec: Artificial real objects}.

\subsection{Dataset on 2015 May}

In this observing run,
eight HSC field-of-views (about 14.2 deg$^2$)
were repeatedly observed with  roughly 1-hr interval
in $g$-band (3--4 visits) and $r$-band (1 visit).
One visit consisted of 3 frames of 2-min exposure
(3 images are co-added for each epoch).

To detect short-timescale transients,
image subtraction was performed by using the  first-visit data as the references.
Therefore, source detection in  difference images
was sensitive to intranight variability.
 From the difference images, 54,672 sources were detected in total.
Features listed in table \ref{table:feature}
were computed for these sources.
The average 5-sigma detection limit was $24.7$ mag.

Furthermore, we performed independent image subtraction,
using the reference data taken on 2014 July 2 and 3 UT.
The difference images between the reference
and this observing run revealed many transients.
Among them, 48 definite transients were 
discovered with visual inspection
and were reported as supernova candidates
by \citet{tominaga15atel1}.
We used the 48 supernova candidates as real transient objects
in section \ref{sec: result sel SN}.

\subsection{Dataset on 2015 Aug}

In this observing run, the weather condition was  poor and 
only 3-hr data were taken under poor seeing of 1.1--1.5 arcsec.
Then, it was not possible to take sufficient baseline
for the image subtraction within the same night.
Hence, we  made difference images
for three HSC field-of-views (5.3 deg$^2$)
with the reference images taken in 2014 Jul.
 From the difference images, 45,019 sources were detected in total.
Features listed in table \ref{table:feature}
were computed for those sources.
The average 5-sigma detection limit was $25.3$ mag.

\subsection{Artificial real objects}
\label{sec: Artificial real objects}

\begin{figure}
 \begin{center}
   \includegraphics[width=8cm, angle=0]{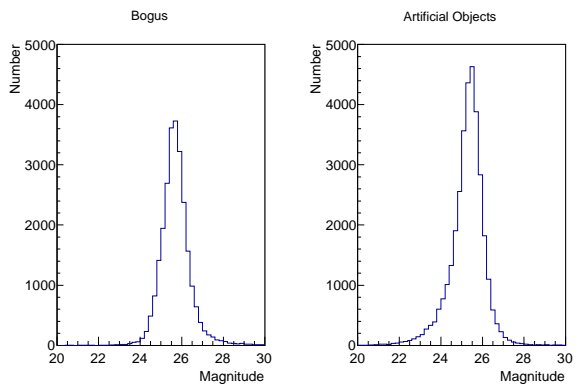}
 \end{center}
\caption{Distribution of magnitudes of bogus objects (left) and
artificial ``real'' objects (right)  in the training data 
made from the May observation.
}\label{fig:mag_dist_artificial_bogus}
\end{figure}

The small number of real sources, as well as 
imbalance between the real and bogus, is a major obstacle
in effective machine-learning.
To improve the performance, we generated 
artificial transient objects to  use
as the real sample  in  training  machines.

 The procedure is as follows.
We constructed spatially-varying PSFs based on the detected stars,
using the HSC pipeline, and 
generated a number of artificial sources from them 
with different brightness  at random positions.
Magnitude ($m$) distribution of artificial sources were  set
to follow  $N(m) = 10^{0.6 m}$, which is expected
when the density of sources with the same
luminosity is constant in the universe.
The faintest magnitude was set  at 27.0 mag, and
1,000 artificial sources were generated
per  CCD chip. These artificial sources were
added to the observed images
to mimic the actual observing conditions. 

We made the training sample containing 
33,742 real and 25,468 bogus sources.
Figure \ref{fig:mag_dist_artificial_bogus}
shows the distribution of magnitudes
 of bogus and artificial ``real'' objects.
This has made up the number of ``real'' objects
comparable to that of the bogus, and accordingly
 enabled training with a sufficient number of ``real'' objects.
The data with artificial sources were reduced and
features were also extracted in the same manner 
as for the real sources.  Note that this strategy was  taken also by \citet{Wright+2015}.

\subsection{Training of Machines}
\subsubsection{AUC Boosting}
\label{AUC Boosting}

\begin{figure*}
 \begin{center}
   \includegraphics[width=6cm, angle=-90]{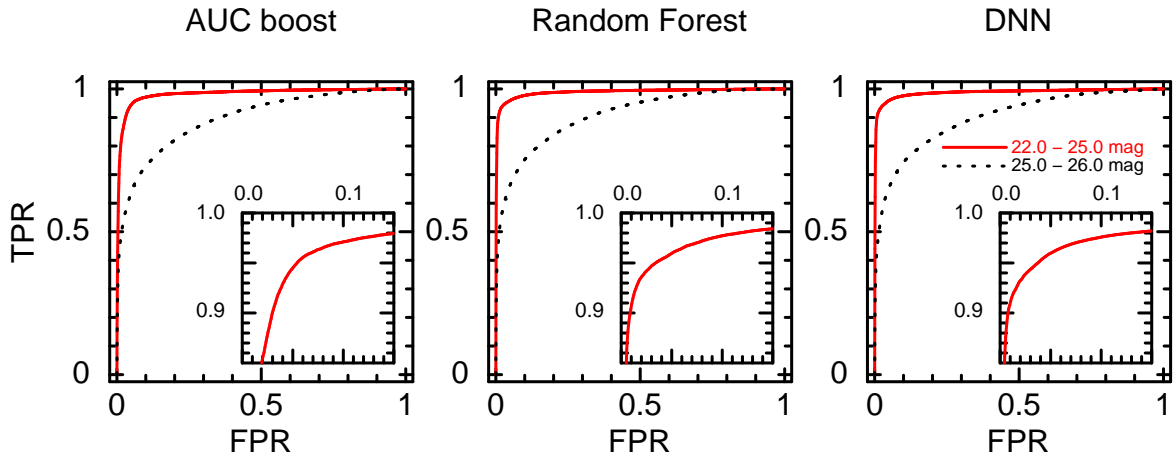}
 \end{center}
\caption{ROC curves of the three machines:
  AUC Boosting (left), Random Forest (middle), and
  Deep Neural Network (right), trained  with the artificial data, 
   based on the May observation.
  Horizontal and vertical axes are FPR and TPR, respectively.
  The red solid and black dotted lines
   show  the curves  for the optical magnitude ranges of  22.0--25.0,
  and  25.0--26.0 mag, respectively.
  Insets show the magnification of the lines  for  22.0--25.0 mag.
}\label{fig:roc3}
\end{figure*}

In the training step, we first determined
the optimal hyper-parameter $\lambda$
by performing cross-validation as follows.
We made 30 partitions of randomly sampled data,
each of which contained the training, validation and test data
with the same size.
For each partition of data, we searched for the optimal $\lambda$
parameter,  evaluating the performance of the machine
with the AUC value for the corresponding validation data.
Next,  fixing the $\lambda$ parameter to the derived optimal,
we trained the machine using the training and validation data,
and evaluated the performance  with the test data.

Figure \ref{fig:roc3} (left\ panel) shows the result: the average of
30 ROC curves  of two magnitude slices for the data of May.
We obtained the average FPR of 3.0\% at the point of
the TPR of 90\% in the magnitude range of 22.0--25.0 mag.

\subsubsection{Random Forest}

We determined the meta parameter $B=100$
by the cross-validation in the same way as 
in the case of AUC boosting (previous subsection). We chose $m_f = 4$, based on the standard 
setting for classification task.
The ROC curves in the magnitude slices are shown
in figure~\ref{fig:roc3}~(middle panel).
We obtained the average FPR of 0.95\% at the point of
the TPR of 90\% for the same range 
as in AUC Boosting.

\subsubsection{Deep Neural Network}
\label{sec:DNN May}

We chose the meta-parameters by preliminary cross validations, as follows.
The number of hidden layers was $3$, and thus our DNN was $5$-layered.
All emission functions of the hidden units were
ReLU (the rectified linear unit).
The emission function of the output layer was the soft-max function.
We used a classical momentum SGD method to optimize the network
 with the mini-batch size $M=4000$.
The number of maximum iterations was $20000$,
but the early stopping rule was employed,
following the standard practice of DNNs.

For the cross-validation,
we split the sample equally in data size into 
training, development, and test dataset,
keeping the ratio between the real and bogus objects,
and prepared $30$ partitions.
For each partition, we performed $30$ repetitions of
training and evaluation to search for the best combination of meta-parameters
 of the number of neural units in hidden layers and the step size of SGD, among exhaustive combinations of them.
Then, we obtained the average test score
of these repetitions.

We obtained the average FPR of 0.85\% at the point of
the TPR of 90\% for the same range as in AUC Boosting.
The ROC curves  are shown in figure \ref{fig:roc3} (right\ panel).

\subsection{Combined machine}

\begin{figure*}
 \begin{center}
   \includegraphics[width=6cm, angle=-90]{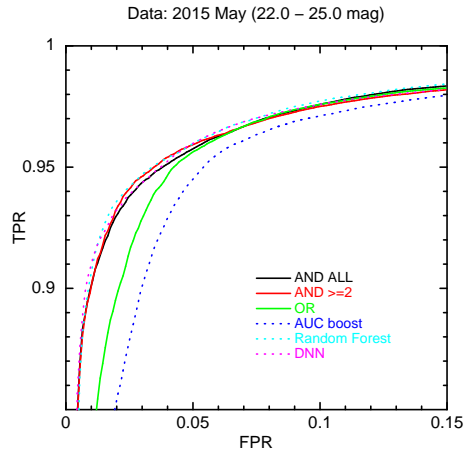}
 \end{center}
\caption{ROC curves of the three ML machines and the combinations of them for the data of 22.0 -- 25.0 mag on 2015 May.
Those of AUC boosting, RF, and DNN are shown in blue, cyan, and magenta dotted curves,
respectively.
Those of ``unanimous voting'', ``majority voting'', and ``safeguard minority opinions''
are shown in black, red, and green solid curves, respectively.
}\label{fig:roc_all}
\end{figure*}

Finally, we combined the results
of the three machines to have a single decision.
There are logically three ways to combine 
three machines  indiscriminately (A, B, and C):  ``unanimous voting'', ``majority voting'',
and ``safeguard minority opinions'',  which are basically ``A \& B \& C'', 
``(A \& B) $|$ (B \& C) $|$ (C \& A)'',
and ``A $|$ B $|$ C'', respectively, where ``\&'' and ``$|$''  denote logical ``AND'' and ``OR'', respectively.
Figure \ref{fig:roc_all} summarizes their ROC curves, together with the ones with individual machines.
We found the FPR of the combined machines at the TPR of 90\%  to be
1.0\%, 1.0\%, and 2.1\% for  the above-mentioned three combinations, respectively.
Among them, the performance of
``safeguard minority opinions'' is not good,
 whereas the performances of ``unanimous voting'' and ``majority voting'' are  good.
We have decided to use ``majority voting'', because
the ``unanimous voting'' would give too tight constraints.

\subsection{Robustness for variation of environment}

\begin{figure*}
 \begin{center}
   \includegraphics[width=6cm, angle=-90]{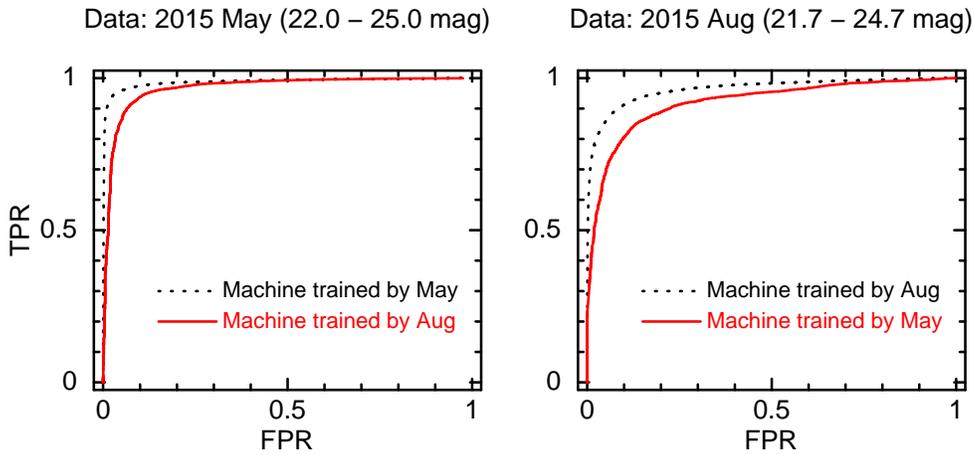}
 \end{center}
\caption{Combined ROC curves as ``majority voting'' of the three ML machines.
Left: ROC curve made by evaluating the data of 22.0 -- 25.0 mag on 2015 May,
using the machine trained  with the same data (black dot) and the data on 2015 Aug (red solid).
Right: ROC curve made by evaluating the data of 21.7 -- 24.7 mag on 2015 August,
using the machine trained  with the same data (black dot) and the data on 2015 May (red solid).
}\label{fig:roc_all_cross}
\end{figure*}

To validate that our machines  are robust against  variations of
environment, we made the following ROC curves.
Figure \ref{fig:roc_all_cross} (left) shows two ROC curves
of the combined ``majority voting'' machine
applied to the artificial data on May;
the black dotted curve shows the result of the machine trained  with the same data,
while the red solid curve is the result of the machine trained  with artificial data on August.
Figure \ref{fig:roc_all_cross} (right) shows the ROC curves of the data  for which May and August are swapped.
In both the panels, the red solid curves are slightly worse than the black dotted curves,
but the differences are small.
We conclude that the machines trained  with the  adopted method  are robust against  changes of
the conditions.

\subsection{Results of selection of supernovae}
\label{sec: result sel SN}

\begin{longtable}{lr}
\caption{Number of selected objects  in the 2015 August data}
\label{table:number of selected objects}
\hline\hline
\multicolumn{1}{c}{Selection} &
\multicolumn{1}{c}{\# of objects} \\
\hline
\endfirsthead
Total         &  45,019 \\ \hline
AUC Boosting  &  21,487 \\
Random Forest &  16,307 \\ 
Deep Neural Network  &   11,645 \\ \hline
Two or three machines
              &  16,888 \\
All machines  &   8,514 \\ \hline
\end{longtable}

Table \ref{table:number of selected objects}  summarizes
the numbers of selected objects obtained  with the three individual 
machines and with the combined  machines, where
 the thresholds are set at the points corresponding to the TPR of 90\%.
 In order not to miss the real objects, we adopted
the ``majority voting'' for the combined machines,  and it reduced the number of objects to $16,888$.

Among 48 supernovae obtained  with the May observation,
26 are in the magnitude range of 22.0 -- 25.0 mag.
We applied the combined machine of ``majority voting'' with
the threshold of TPR of 90\% at this magnitude range.
Then, 22 sources, or 85\% of them, were  accepted.

\subsection{Quick selection of supernovae}
\label{sec: quick sel SN}

Before the August observation, we had installed the trained machines
in order to run transient search  immediately after the observation 
(see section \ref{sec: result sel SN}).
We then visually inspected every object detected, selected ten clean sources as  definite candidates of supernovae,
and reported the list to Astronomers' Telegram \citep{tominaga15atel2}.
This  whole process was  carried out within the same day as the observation.

\section{Discussion and Conclusion}
\label{sec: Discussion and Conclusion}

\begin{figure*}
 \begin{center}
   \includegraphics[width=16cm, angle=0]{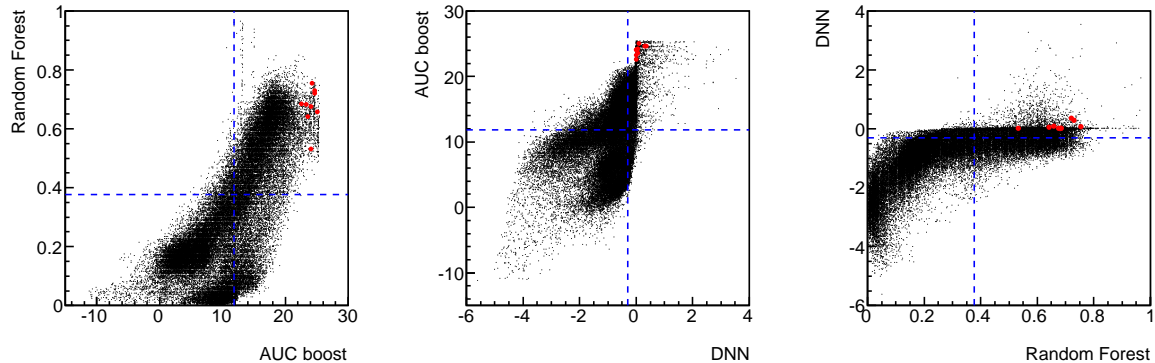}
 \end{center}
\caption{Cross-relations of scores obtained  with the three machines
  for  the entire sample (black) and the selected sample of
  ten candidates of supernovae 
  reported to Astronomers' Telegram \citep{tominaga15atel2} (red)
  for the observation data on 2015 August\ 19.
  The left,  middle, and right panels  plot
  those of Random Forest v.s. AUC Boosting,
  AUC Boosting v.s. Deep Neural Network,
  and Deep Neural Network v.s. Random Forest, respectively.
  Horizontal and vertical dashed lines show the thresholds
  corresponding to the TPR of 90\%.}
\label{fig:score_scat3_sel10}
\end{figure*}

We performed Subaru/HSC observations in 2015 May and August, made transient searches,
and  applied machine-learning techniques to the result to reduce the
bogus transient objects.
We  have developed real-bogus classifiers as the core function for it, using
the three machine-learning methods of AUC Boosting,
Random Forests, and Deep Neural Network, and
 then made the combined classifier as  their majority voting.
We have installed  our machines
in the analysis pipeline of the HSC,
and  successfully found real supernovae within the same day as
the observation, demonstrating the power of our method.
Now, we have  completed the preparation for 
the forthcoming HSC/SSP transient survey observation.

In training our machines, we used artificial objects,
because  the data are  highly imbalanced between real/bogus objects,
and it was found to be crucial to make
good machines efficiently.
Although the HSC survey data are technically more difficult to deal with than other survey data,  given that the HSC
survey is deeper than other surveys, we have achieved  the results  comparable to  the similar studies in other surveys.

The cross-relations between three machines
in figure \ref{fig:score_scat3_sel10} show that 
 none of the three is significantly better than the other two.
Therefore, combining the machines is beneficial.
We used a moderate selection with all the combinations
of multiple machines, ``majority voting'',
to avoid missing some real objects.
For the machine, robustness against variation of environment was confirmed.

In this paper we have focused on real-bogus separation with 
machine-learning methods,
and  have demonstrated that they were indeed  useful.  Their use in extracting scientific results
out of Big-data in astronomy is promising.
As the next step, we will use machine-learning  for classification of types of transients,
by combining timing and color information,
as well as the shape of objects.

\begin{ack}
We thank  the Subaru Hyper Suprime-Cam team.
This work is supported by Core Research for Evolutionary Science
and Technology (CREST), Japan Science and Technology Agency (JST).
It is also supported by the research grant program
of Toyota foundation (D11-R-0830)
and was in part supported by
Grants-in-Aid for Scientific Research of JSPS (15H02075),
MEXT (15H00788),
and the World Premier International Research Center Initiative, MEXT, Japan.
This paper makes use of software developed for the LSST.
We thank the LSST Project for  allowing their code available
as free software at http://dm.lsstcorp.org.
\end{ack}

\appendix

\end{document}